\documentclass[prl,aps,twocolumn,superscriptaddress,showpacs,preprintnumbers,asmath,amssymb]{revtex4-1}

\usepackage{color}
\usepackage{graphicx}
\usepackage{ulem}
\usepackage{float}
\usepackage{here}
\usepackage{bm}
\usepackage{amstext}
\usepackage{amsfonts}
\usepackage{dcolumn} 
\usepackage[colorlinks=true,urlcolor={blue},citecolor={blue}]{hyperref}

\def\textbf#1{\boldsymbol{#1}}

\begin{document}

\title{Pressure-induced phase transition and superconductivity in YBa$_{2}$Cu$_{4}$O$_{8}$}
\author{S. M. Souliou}
\affiliation{Max-Planck-Institut f\"{u}r Festk\"{o}rperforschung, Heisenbergstrasse 1, D-70569 Stuttgart, Deutschland}
\author{A. Subedi}
\affiliation{Centre de Physique Th\'{e}orique, \'{E}cole Polytechnique, CNRS, 91128 Palaiseau Cedex, France}
\author{Y.T. Song}
\affiliation{Max-Planck-Institut f\"{u}r Festk\"{o}rperforschung, Heisenbergstrasse 1, D-70569 Stuttgart, Deutschland}
\affiliation{Research \& development Center for Functional Crystals, Beijing National Laboratory for Condensed Matter Physics, Institute of Physics, Chinese Academy of Sciences, Beijing 100190, China}
\author{C.T. Lin}
\affiliation{Max-Planck-Institut f\"{u}r Festk\"{o}rperforschung, Heisenbergstrasse 1, D-70569 Stuttgart, Deutschland}
\author{K. Syassen}
\affiliation{Max-Planck-Institut f\"{u}r Festk\"{o}rperforschung, Heisenbergstrasse 1, D-70569 Stuttgart, Deutschland}
\author{B. Keimer}
\affiliation{Max-Planck-Institut f\"{u}r Festk\"{o}rperforschung, Heisenbergstrasse 1, D-70569 Stuttgart, Deutschland}
\author{M. Le Tacon}
\affiliation{Max-Planck-Institut f\"{u}r Festk\"{o}rperforschung, Heisenbergstrasse 1, D-70569 Stuttgart, Deutschland}

\date{\today}

\begin{abstract}We investigate the pressure and temperature dependence of the lattice dynamics of the underdoped, stoichiometric, high temperature superconductor YBa$_{2}$Cu$_{4}$O$_{8}$ by means of Raman spectroscopy and \textit{ab initio} calculations. This system undergoes a reversible pressure-induced structural phase transition around 10 GPa to a collapsed orthorhombic structure, that is well reproduced by the calculation. The coupling of the B$_{1g}$-like buckling phonon mode to the electronic continuum is used to probe superconductivity. In the low pressure phase, self-energy effects through the superconducting transition renormalize this phonon, and the amplitude of this renormalization strongly increases with pressure. Whereas our calculation indicates that this mode's coupling to the electronic system is only marginally affected by the structural phase transition, the aforementioned renormalization is completely suppressed in the high pressure phase, demonstrating that under hydrostatic pressures higher than 10 GPa, superconductivity in YBa$_{2}$Cu$_{4}$O$_{8}$ is greatly weakened or obliterated.
\end{abstract}

\pacs{74.25.nd, 74.72.Gh, 71.15.Mb, 71.18.+y, 74.62.Fj}


\maketitle

High pressure constitutes a valuable tool for the study of complex phase diagrams such as the ones encountered in high temperature copper oxide superconductors (cuprates)~\cite{Schilling2007}, where various competing phases can coexist. In addition to insights about the mechanism of superconductivity in cuprates, high pressure studies have also led to the highest superconducting critical temperatures \textit{T$_{c}$} ever reported~\cite{Chu1993,Gao1994}.
The pressure dependence of \textit{T$_{c}$} is affected by several parameters, including the pressure-induced hole transfer towards the CuO$_{2}$ planes and structural parameters such as the buckling angle of CuO$_{2}$ planes or the interplanar distance~\cite{Chen2000,Neumeier1993,Murayama1991,Dabrowski1996}. In the widely studied YBa$_{2}$Cu$_{3}$O$_{6+x}$ family the pressure dependence of the superconducting properties gets further complicated by the occurrence of oxygen relaxation effects in partially filled Cu-O chains~\cite{Sadewasser2000}.

\begin{figure}
\includegraphics[width=1\linewidth]{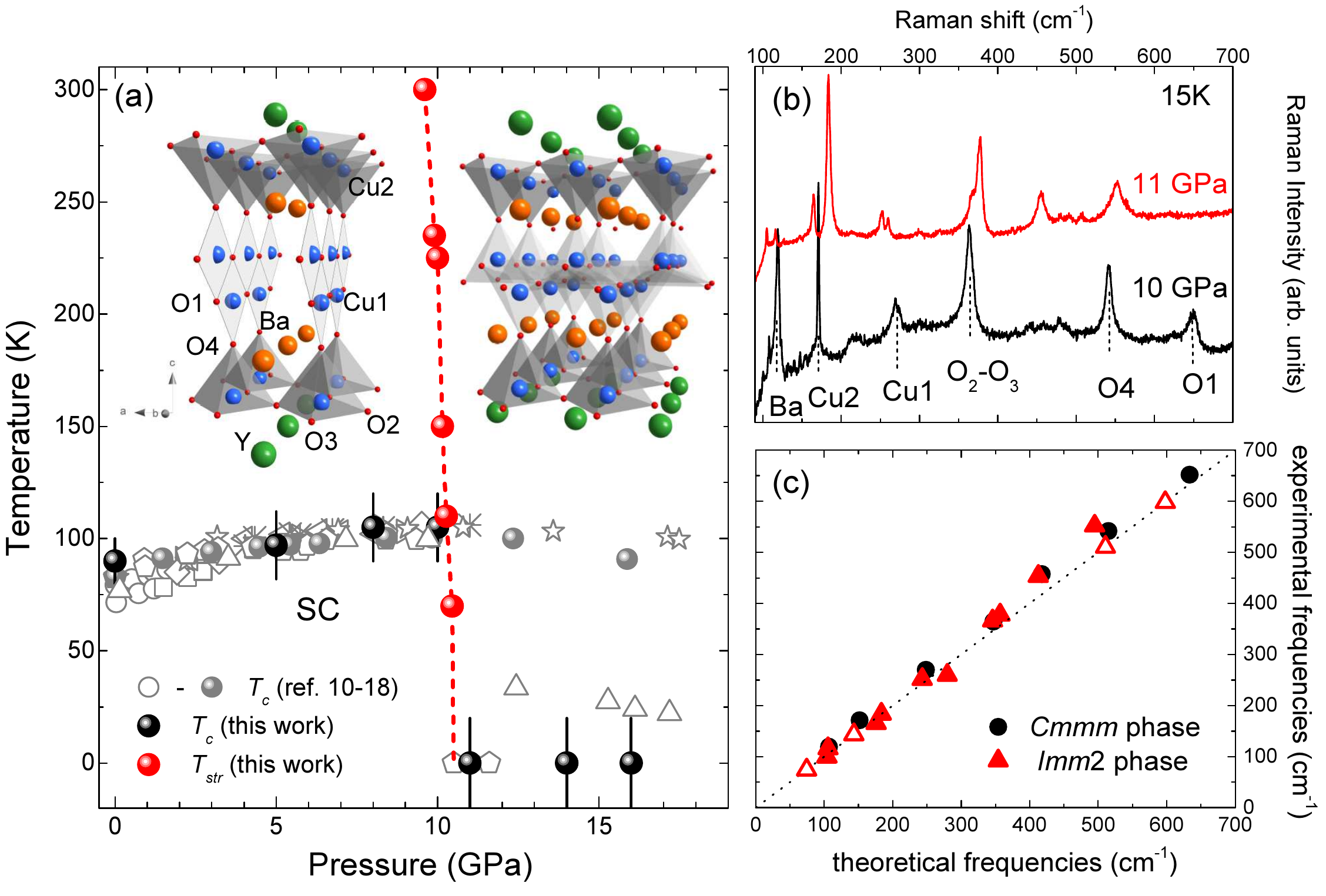}
\caption{(a) (P,T) phase diagram of YBa$_{2}$Cu$_{4}$O$_{8}$. The red (black) symbols show the structural (superconducting) transition temperature \textit{T$_{str}$} (\textit{T$_{c}$}) as derived from our Raman measurements and the grey symbols are taken from resistivity and susceptibility data of references~\cite{Bucher1989,Diederichs1991,Tissen1991a,Braithwaite1991,Mori1991,Vaneenige1990,Scholtz1992,Mito2014,Nakayama2014}. The illustrated structures of the two phases are based on the data of reference~\cite{Su2007}. (b) Raman spectra of YBa$_{2}$Cu$_{4}$O$_{8}$ recorded at 15 K under a pressure of 10 and 11 GPa. (c) Experimental and calculated Raman phonon frequencies in the two phases. The experimental frequencies correspond to the spectra of panel (b). Open symbols correspond to modes predicted by the theory that we do not observe.} \label{Fig1}
\end{figure}

These complications are absent in YBa$_{2}$Cu$_{4}$O$_{8}$ (Y124), a stoichiometric, underdoped cuprate with \textit{T$_{c}$} $\sim$ 80 K, thanks to the filled double Cu-O chains of its structure~\cite{Karpinski1988}. High pressure transport and susceptibility measurements have revealed an initially large pressure derivative of \textit{T$_{c}$} ($\sim$ 5 K/GPa)~\cite{Bucher1989,Diederichs1991,Tissen1991a,Braithwaite1991,Mori1991,Vaneenige1990,Scholtz1992,Mito2014,Nakayama2014}, leading to a maximum value of $\sim$ 105 K at 10 GPa (Fig.~\ref{Fig1}-a).
At higher pressures, the situation is more controversial, and depending on the pressure conditions, \textit{T$_c$} is reported either to decrease~\cite{Vaneenige1990,Scholtz1992,Mito2014,Nakayama2014} or to disappear~\cite{Tissen1991a,Mito2014}.
Interestingly, x-ray diffraction (XRD) studies performed at room temperature revealed the occurrence of a first-order, pressure-induced, structural phase transition around $\sim$ 10 GPa~\cite{Su2007,Wang2007,Nakayama2014}. The structural refinement showed that the CuO$_2$ planes remain almost unaffected by the transition, whereas the double Cu-O chains tilt, resulting in a collapsed orthorhombic structure. Based on the XRD extinction rules, a non-centrosymmetric (\textit{Imm}2)~\cite{Su2007,Wang2007} and a centrosymmetric (\textit{Immm})~\cite{Nakayama2014} space group were proposed for the high pressure structure. Since there is only a limited number of non-centrosymmetric superconductors, all with rather modest \textit{$T_c$} but highly unconventional properties, such as very large, non-Pauli limited, upper critical fields~\cite{Bauer2011}, it is worth exploring whether the high pressure phase of Y124 is actually superconducting.

Inelastic light (Raman) scattering is particularly suitable to address this issue given that, on the one hand, some of the Raman active phonons are coupled to the electronic continuum and strongly renormalize across the superconducting transition and that, on the other hand, the structural transition is clearly reflected in the high pressure Raman spectra of Y124 with the appearance of several new modes~\cite{Su2007}.
Furthermore, unlike many other techniques, the high pressure Raman measurements can be performed using He as a pressure transmitting medium, ensuring the best possible hydrostatic conditions.
We have carried out a detailed study of the effects of temperature and pressure (up to 16 GPa) on the lattice dynamics of Y124, that allowed us to map out the (P,T) phase diagram of this compound. We show that, in the low pressure phase, the well-known superconductivity induced renormalization of the planar Cu-O bond-bending ("buckling") mode is strongly enhanced as pressure is increased. Above $\sim$ 10 GPa, on the other hand, after the structural phase transition occurs no renormalization of this mode is observed. The \textit{ab initio} calculations, that accurately account for the structure and phonon spectra of both phases, indicate that the coupling of this mode to the electronic background is only weakly affected by the structural transition. This leads us to infer that superconductivity is either absent or greatly weakened in the high pressure phase of Y124.

High quality untwinned Y124 single crystals were grown at ambient air pressure by a flux method as described elsewhere~\cite{Song2007,Sun2008}.
The Raman measurements have been performed on a macro Raman setup (Jobin-Yvon, T64000) equipped with a liquid-nitrogen-cooled charge coupled device (CCD). The 647.1 nm line of a Krypton laser was used as excitation. The incident photon propagation direction was aligned along the crystallographic \textit{c}-axis of the sample (backscattering geometry), selecting $c-$axis polarized modes with the A$_{g}$ symmetry in the low pressure phase.
High pressures were applied using a gasketed Stuttgart type diamond anvil cell, loaded with condensed He as a pressure transmitting medium. Pressure was calibrated with the ruby luminescence method~\cite{Syassen2008}. To check hydrostaticity, the pressure was systematically checked on several ruby spheres positioned around the studied sample in the gasket hole~\cite{Klotz2009}. Additionally, pressure was always changed at temperatures above the helium melting line (below 11 GPa) in order to avoid residual stresses acting on the sample. For the low temperature measurements the cell was immersed in a helium bath cryostat.

Figure~\ref{Fig1}-b shows low temperature ($T = $15 K) unpolarized Raman spectra of Y124 recorded at 10 and 11 GPa. As expected from group theory, and in agreement with previous reports~\cite{Heyen1990,Heyen1991}, 7 A$_{g}$ symmetry phonons are observed in the 10 GPa spectrum. These modes correspond to vibrations along the \textit{c}-axis of Ba (104 cm$^{-1}$), plane Cu2 (151 cm$^{-1}$), chain Cu1 (257 cm$^{-1}$), plane O2-O3 out-of-phase (the so-called B$_{1g}$-like buckling mode, 338 cm$^{-1}$), plane O2+O3 in-phase (431 cm$^{-1}$), apical O4 (496 cm$^{-1}$), and chain O1 (600 cm$^{-1}$) atoms (the frequencies are given at 10 GPa, 15K). The pressure-induced hardening of the Raman frequencies is quantitatively in line with earlier reports~\cite{Lampakis2005}. In our experimental geometry and within the applied pressure step ($\sim$ 1.5 GPa) we did not observe any deviations from linearity in the pressure dependence of the O2+O3 and Cu2 phonon modes frequencies and no splitting of the in-phase O2+O3 mode, like those reported in previous studies~\cite{Boley1994,Watanabe19949226,Kakihana1994,Osada2001142,Lampakis2005}.

At 11 GPa the low temperature Raman spectrum is drastically altered, evidencing the transition to the new structural phase. The overall Raman spectrum of the new phase is similar to the one at room temperature~\cite{Su2007}.
In addition to the "descendants" of the O2-O3, O2+O3 and O4 phonon modes of the initial phase, several new phonon modes appear around 100, 110, 155 and 175 cm$^{-1}$ at high pressure.
Qualitatively, this is not surprising given that the CuO$_2$ planes remain almost unaffected by the structural change and retain their initial topology.
Contrary to this, the O1 and Cu1 chain-related modes are largely suppressed in the new phase, in accordance with the collapse of the double chains suggested by the XRD data.
Note that downstroke measurements (not shown here) indicate that the structural transition is reversible, without any noticeable hysteresis effects. Moreover, the quality of the spectra in the downstroke measurements is comparable to that of the upstroke measurements, indicating that the crystal is not damaged throughout the entire high pressure experiment.

Based on group theory considerations~\cite{Kroumova2003}, the symmetries and the number of the expected zone center modes in the \textit{Imm}2 and \textit{Immm} phases are:
\begin{equation}\label{eq1}
\Gamma(\textit{Imm}2)=15A_1+7A_2+15B_1+8B_2
\end{equation}
and
\begin{equation}\label{eq2}
\Gamma(\textit{Immm})=7A_g+7B_{2g}+7B_{3g}
\end{equation}
According to the selection rules, A$_1$,A$_2$ and A$_{g}$ symmetry modes can be probed in our experimental geometry. Complementary infra-red measurements could provide further hints regarding the non-centrosymmetry, but already here, given that in the low temperature Raman spectra of the high pressure phase we observe 10 phonon modes (Figure~\ref{Fig1}-b), the \textit{Imm}2 space group appears more likely.
To gain more insights about the new phonon modes, we have calculated the phonon frequencies and electron-phonon couplings using density functional theory (DFT) calculations within the local density approximation (LDA). Details are given in the Supplemental Material section~\cite{Supplementary} and in refs.~\cite{Blochl1994, Kresse1999, Kresse1996, Baroni2001, Giannozzi2009} therein. These calculations were done using the experimental lattice parameters at hydrostatic pressures of 10 ($Cmmm$ phase with a volume of 188.32 \AA$^3$) and 11 GPa ($Imm2$ phase with a volume of 177.10 \AA$^3$)~\cite{Su2007}, but with relaxed internal parameters. As can be seen from Figure~\ref{Fig1}-c, we find excellent agreement between the calculated and experimental frequencies for these modes.

\begin{figure}
\includegraphics[width=1\linewidth]{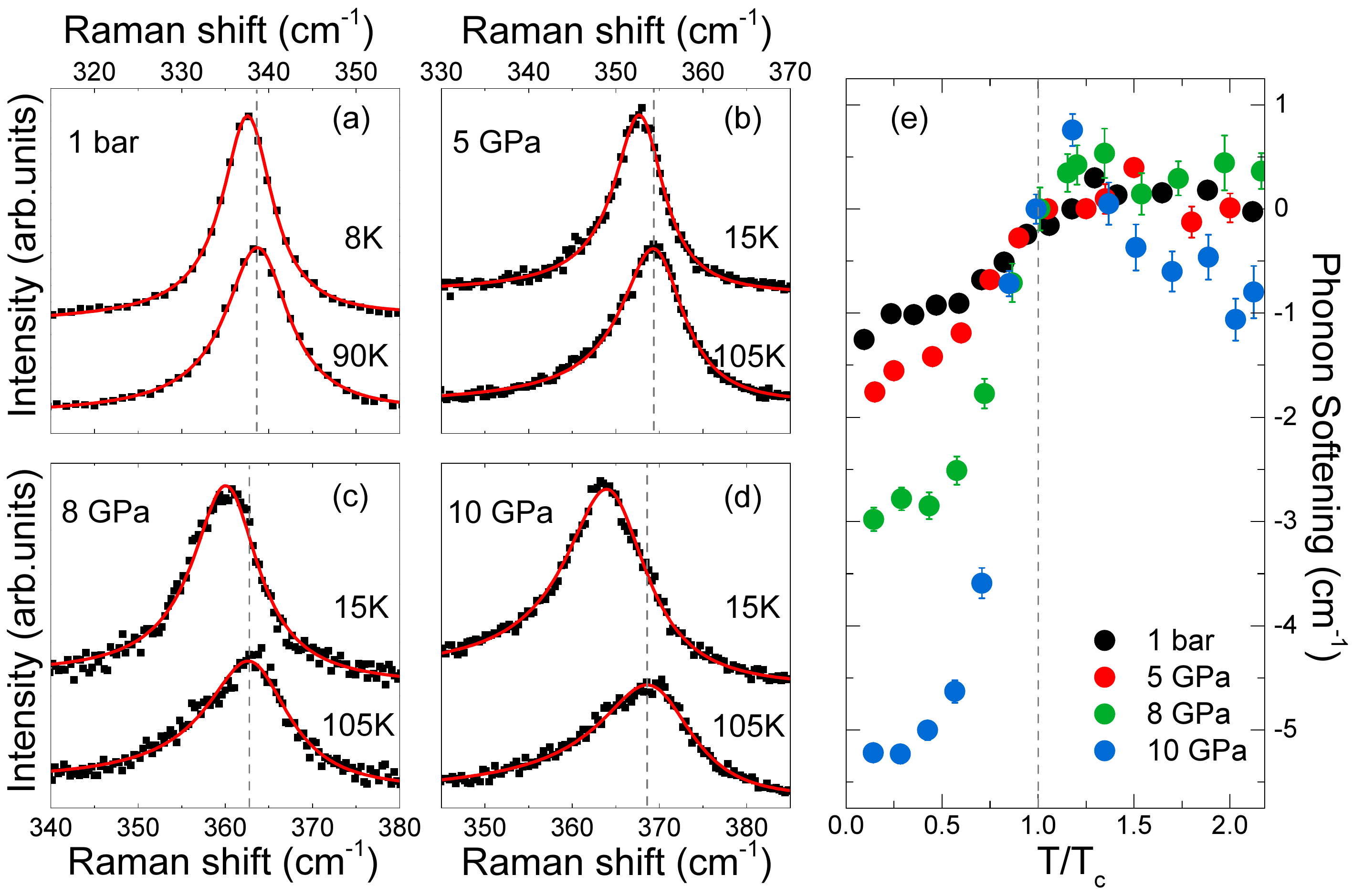}
\caption{(a-d) Raman frequency region of the O2-O3 phonon at $\sim$ \textit{T$_{c}$} and deep in the superconducting state and (e) phonon softening below \textit{T$_{c}$} at ambient pressure, 5, 8 and 10 GPa.} \label{Fig2}
\end{figure}

We now turn to the effect of temperature on the phonon spectra. At ambient pressure, several $A_g$ modes renormalize across the superconducting transition. The effect is particularly clear on the Ba~\cite{Heyen1990} and on the O2-O3 B$_{1g}$-like buckling phonon mode and has been widely studied in Y124 and Y123 systems~\cite{Altendorf1993,Bakr2009,Cardona1999,Friedl1990,Limonov2000}.
The coupling of a phonon to the underlying electronic system results in an asymmetric 'Fano' lineshape, which is significantly renormalized below \textit{T$_{c}$} as a result of the changes in the phonon self-energy as the gap in the electronic continuum opens~\cite{Nicol1993,Zeyher1990}.

In Figs.~\ref{Fig2}-a to d, we take a closer look at the buckling phonon mode at ambient pressure, 5, 8 and 10 GPa just above the superconducting transition, and at base temperature.
At ambient pressure, the phonon softens by $\sim$ 1 cm$^{-1}$ between $T_c$ and 8 K, in good agreement with previous reports~\cite{Heyen1991}. Upon pressure increase, the superconductivity induced softening increases strongly to become as large as $\sim$ 5 cm$^{-1}$ at 10 GPa. This can be seen in Fig.~\ref{Fig2}-e, where we have plotted the amplitude of the phonon softening as a function of the reduced temperature $T/T_c$ (defining $T_c$ as the onset of the softening yields good agreement with published transport data, as seen in Fig.~\ref{Fig1}-a). 

Qualitatively, the evolution of the phonon renormalization in Y124 under high pressure is reminiscent of the effect of hole doping in Y123~\cite{Altendorf1993,Limonov2000,Friedl1990}, where the phonon renormalization is weak in the underdoped regime and becomes larger (with amplitude comparable to the one of Y124 at 10 GPa) at optimal doping where $T_c$ is maximized.
This frequency shift directly relates to the real part of the phonon self-energy by
\begin{equation}\label{eq3}
\frac{\Delta\omega}{\omega_{0}}=\frac{1}{N_F}\lambda_{ph}Re\Pi(\omega_{0})
\end{equation}
where $N_F$ is the the electronic density of states at the Fermi level, $\lambda_{ph}$ is the total electron-phonon coupling constant and Re$\Pi$ is the real part of the phonon self-energy.

In Y123, the doping increase of the renormalization is attributed to $Re\Pi(\omega_{0})$, that is, in the $d-$wave case, maximized as the phonon frequency $\omega_{0}$ matches the amplitude of the superconducting gap 2$\Delta$.
The pressure-induced hardening of the mode frequency (as discussed in the data of reference \cite{Goncharov1994} for similar observations in underdoped YBa$_{2}$Cu$_{3}$O$_{6+x}$ under pressure) combined with pressure-induced changes of 2$\Delta$~\cite{Zhou1996669} could therefore account for our observation.
We can however not rule out a contribution of a pressure-induced increase of $\lambda_{ph}$, supported by the increase of the normal state phonon lineshape asymmetry with pressure (Figure~\ref{Fig2}). In any event, these data demonstrate that $Cmmm$ Y124 remains superconducting up to 10 GPa.


\begin{figure}
\includegraphics[width=0.8\linewidth]{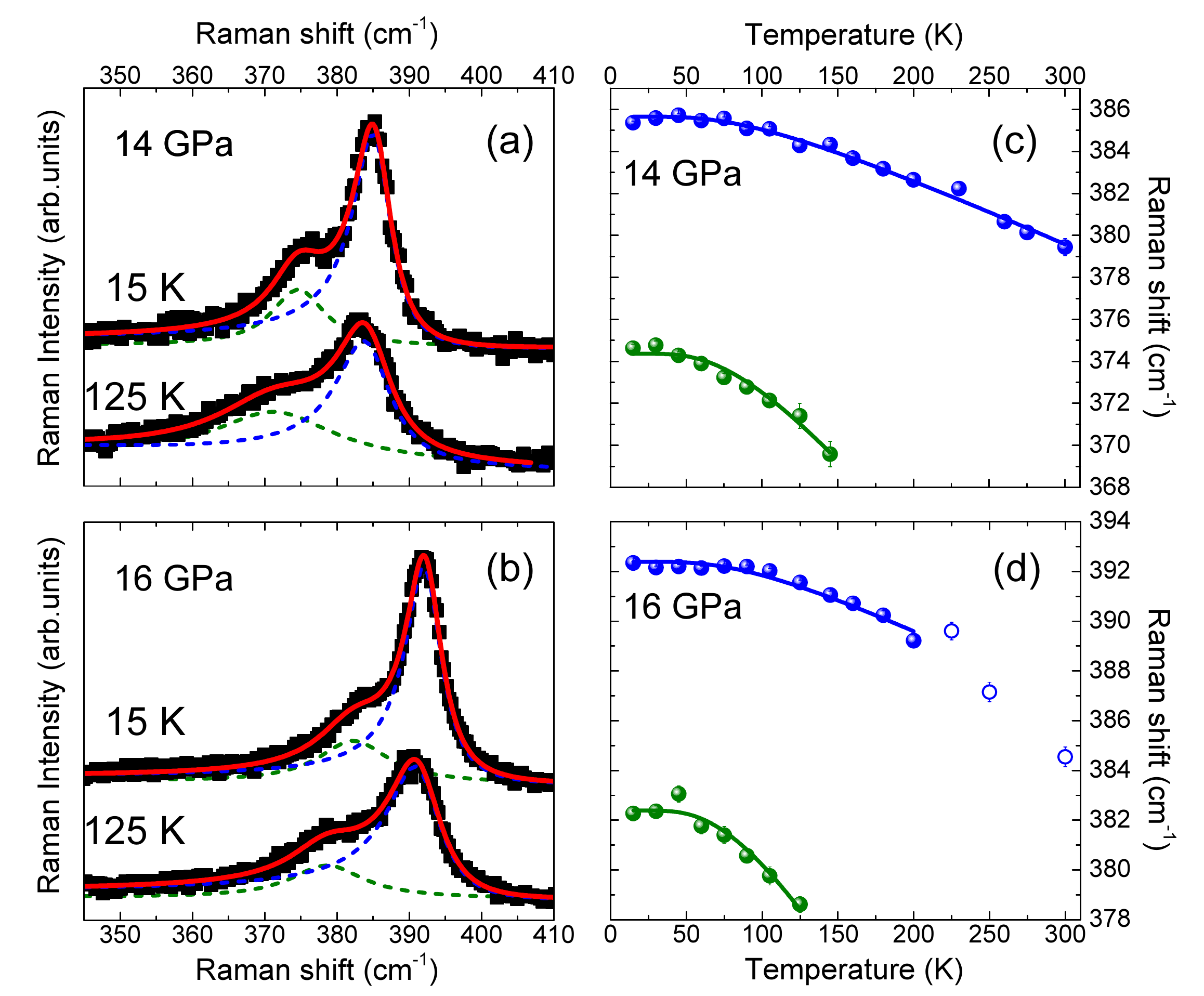}
\caption{(a, c) Raman frequency region of the O2-O3 mode in the new phase at 125 and 15 K and (b, d) temperature dependence of the observed phonons at 14 and 16 GPa. The solid lines are fits based on the anaharmonic decay of optical phonons model~\cite{Klemens1966}.} \label{Fig3}
\end{figure}

Above this pressure, the behavior of the buckling phonon at low temperatures becomes very different. In Fig.~\ref{Fig3}, we show Raman spectra at 14 and 16 GPa recorded at 125 K and 15 K in the frequency region where the buckling mode (that has now the $A_1$ symmetry) is seen in the high pressure phase.
At room temperature, it appears as a single broad asymmetric peak~\cite{Su2007}. Cooling however reveals two lines around $\sim$ 380 and $\sim$ 390 cm$^{-1}$ that can unambiguously be resolved as they sharpen below 150 K (for T$>$150 K, the low-energy mode frequency can not be accurately determined). Based on our calculations, we assign the lower energy one to the buckling mode, and the higher energy one to in-plane vibrations of plane oxygen atoms.
Note that the two modes exhibit different slopes in their temperature dependence, ruling out the possibility that the presence of two peaks could be due to non-hydrostatic conditions.

The main result is that, at variance with the behavior reported at lower pressure, both modes continuously harden down to the lowest measured temperature, as expected from an anharmonic decay model~\cite{Klemens1966} (Figure~\ref{Fig3}). No softening is observed in the temperature region where the transport and magnetic susceptibility experiments have reported signatures of a superconducting transition~\cite{Vaneenige1990,Scholtz1992,Mito2014}.


\begin{figure}
\includegraphics[width=0.8\linewidth]{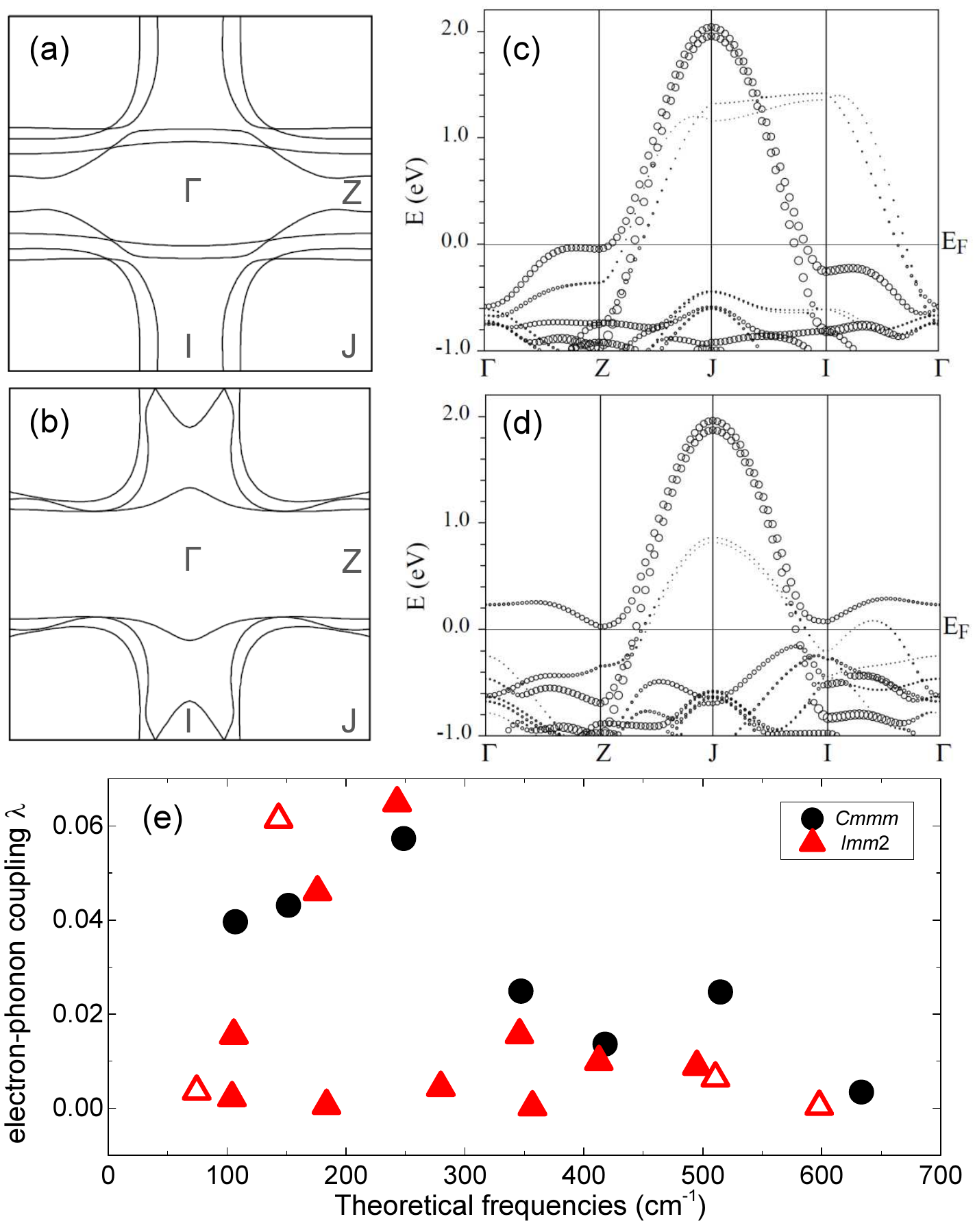}
\caption{(a,b) Fermi surface of YBa$_{2}$Cu$_{4}$O$_{8}$ viewed along the k$_z$ direction in the (a) original and (b) high pressure phase. (c,d) Band structure along selected symmetry directions in the basal plane in the (c) original and (d) high pressure phase. The circle size reflects the band character (big-Cu plane, small-Cu chain). (e) Calculated electron-phonon coupling constants in the two phases. Open symbols correspond to modes predicted by the theory that are not observed in the Raman spectra.} \label{Fig4}
\end{figure}

In addition to the absence of renormalization effects in the two modes of Figure~\ref{Fig3}, no anomalies have been observed for any of the phonon modes of the new structural phase, unlike the many reported for those of the $Cmmm$ phase~\cite{Heyen1990}.
According to our previous discussion, the absence of phonon renormalizations in the high pressure phase could either be attributed to a suppression of $\lambda_{ph}$ or simply to the absence of superconductivity. To test the first hypothesis, following the crystal structure and lattice dynamical calculations presented above, we have calculated the electronic structure and the electron-phonon coupling constant for each zone center phonon in both phases.

The calculated Fermi surface in the $Cmmm$ phase is shown in Figure~\ref{Fig4}-a and c and is in good agreement with previous reports~\cite{Bangura2008,Yu1991_2}. It consists of two hole like quasi-2D cylinders with mostly planar character centered on the Brillouin zone corner \textit{J}=($\pi$,$\pi$), and two 1D open sheets of mostly chain character running along the k$_x$ direction.

In the high pressure phase, the Fermi surface retains its major characteristics exhibiting again two closed sheets centered on ($\pi$,$\pi$) and a distorted open sheet running along along the k$_x$ direction. In this case though, the initial outer antibonding plane band is lifted and no longer crosses the Fermi level. The inner and outer sheets of the Fermi surface have now a chain character (due to the collapse of the double Cu-O chains structure, chain bands that were initially 1D become 2D in the new structure), while the middle sheet is formed by the bonding band of the CuO$_2$ planes.

The electron-phonon linewidth $\gamma^{\mathbf{q},\nu}_{ph}$ and coupling $\lambda^{\mathbf{q},\nu}_{ph}$ for each phonon branch can be calculated using
\begin{equation}\label{eq4}
\gamma^{\mathbf{q},\nu}_{ph} = \frac{2\pi}{N_k} \sum_{\mathbf{k},n,m}
|g_{\mathbf{k}n,\mathbf{k+q}m}^{\nu}|^{2} \delta(\epsilon_{\mathbf{k}n})\delta(\epsilon_{\mathbf{k+q}m}) \\
\end{equation}
and
\begin{equation}\label{eq5}
\lambda^{\mathbf{q},\nu}_{ph} = \frac{1}{2 \pi N_F}\frac{\gamma_{ph}}{\omega^2_{\mathbf{q}\nu}} \\
\end{equation}
where $N_k$ is the number of $k$ points in the sum, $\epsilon_{\mathbf{k}n}$ is the electronic energy at wave vector $\mathbf{k}$ and band index $n$ measured with respect to the Fermi level, $|g_{\mathbf{k},\mathbf{k+q}}^{\nu,n,m}|^{2}$ is the matrix element for an electron in the state $|n\mathbf{k}\rangle$ scattering to $|m\mathbf{k+q}\rangle$ through a phonon $\omega _{\mathbf{q}\nu}$ with wave vector $\mathbf{q}$ and branch index $\nu$.

The calculated values for the electron-phonon coupling of the zone center (${\bf q}$ ={\bf 0}), Raman active phonons of Y124 have been evaluated in the two phases, and the results are reported on Fig.~\ref{Fig4}-e. In the new phase, we find that the electron-phonon coupling constants of those of the $A_1$ modes that correspond to the $A_g$ modes of the $Cmmm$ phase do not vanish and remain comparable to the ones of the $Cmmm$ phase. In particular, the coupling constant of the buckling mode is reduced by about $\sim$ 35\%, which would lead to a sizeable superconductivity-induced renormalization if, as suggested by transport and magnetic susceptibility experiments ~\cite{Vaneenige1990,Scholtz1992,Mito2014}, $T_c$ (and therefore 2$\Delta$) were only marginally affected by the structural phase transition.
We do not observe systematic anomalies in the whole temperature range explored for any of the observed phonons. This is in particular the case also for the Ba or the Cu2 mode around 105 and 160 cm$^{-1}$ (see Supplemental Material~\cite{Supplementary}). This suggests that, possibly due to the drastic changes of the electronic structure induced by the double chain collapse reported above, under hydrostatic conditions, the high pressure phase of Y124 is not superconducting. However, our Raman data can not completely rule out the possibility of a superconducting ground state with a much lower $T_{c}$ (and 2$\Delta$), bounded to about $\sim$ 30 K.

In conclusion, we have mapped out the pressure and temperature phase diagram of the stoichiometric, underdoped, Y124 high temperature superconductor.
The increase of $T_c$ under hydrostatic pressure is accompanied by an enhancement of the superconductivity-induced renormalization of the buckling mode that persists up to $\sim$ 10 GPa. Above this pressure, a structural phase transition to a collapsed orthorhombic structure, well captured by first principle calculations, occurs and we do not observe any phonon anomalies at low temperatures. Further exploration of this new metallic (or weakly superconducting) underdoped cuprate system may provide fresh insights into the electronic structure conducive to high-temperature superconductivity.

\section{Acknowledgement}
This work was supported by the Swiss National Supercomputing Centre (CSCS) under project ID s404 and by the European Commission under the Seventh Framework Programme:
Research Infrastructures (grant agreement no 226716) and the European project SOPRANO under Marie Curie actions (grant no PITNGA-2008-214040).


\end{document}